\documentclass[runningheads]{llncs}
\usepackage[T1]{fontenc}
\usepackage{graphicx}

\usepackage{booktabs}

\usepackage{multirow}  
\usepackage{tabularx}
\usepackage[hidelinks]{hyperref}

\begin{document}
%
\title{Detecting Student Intent for Chat-Based Intelligent Tutoring Systems}
%
%
\author{Ella Cutler\inst{1}\orcidID{0009-0001-2661-142X} \and
Zachary Levonian\inst{2}\orcidID{0000-0002-8932-1489} \and
S. Thomas Christie\inst{2}\orcidID{0009-0006-2900-4939}}
%
\authorrunning{E. Cutler et al.}

%
\institute{Math Academy\\
\and
Digital Harbor Foundation\\
}
%
\maketitle              
\begin{abstract}
Chat interfaces for intelligent tutoring systems (ITSs) enable interactivity and flexibility.
However, when students interact with chat interfaces, they expect dialogue-driven navigation from the system and can express frustration and disinterest if this is not provided.
Intent detection systems help students navigate within an ITS, but detecting students' intent during open-ended dialogue is challenging.
We designed an intent detection system in a chatbot ITS, classifying a student's intent between continuing the current lesson or switching to a new lesson.
We explore the utility of four machine learning approaches for this task---including both conventional classification approaches and fine-tuned large language models---finding that using an intent classifier introduces trade-offs around implementation cost, accuracy, and prediction time.
We argue that implementing intent detection in chat interfaces can reduce frustration and support student learning.

\keywords{classification \and intelligent tutoring systems \and large language models \and intent detection.}
\end{abstract}
%
%
%


\section{Introduction}


Some intelligent tutoring systems (ITSs) use chat interfaces to allow for flexible student inputs.
Large language models (LLMs) are increasingly integrated within ITSs as chatbots that process student's text inputs to generate conversational responses~\cite{wu_improving_2025,levonian_designing_2025}.
LLMs are appealing for their potential to generate responsive, contextually-relevant output this is grounded in pedagogical goals.
However, integrating LLMs into ITSs is not easy: ITSs need to track and update internal state in order to provide safe, structured pedagogical experiences~\cite{vanlehn_behavior_2006}.
Chat interfaces carry with them an expectation of dialogue-driven navigation, and students can become disinterested or frustrated if their messages fail to update the ITS's internal state appropriately.
Our research team observed many frustrated messages from students using the conversational lesson feature we investigate in this study: students expected to be able to change the current task with minimal context, prompting us to develop a classifier to determine when a user wants to change lessons. This classification task is called \textit{intent detection}.

Intent detection is the problem of determining a user's intent from their  messages to a system and is required for tracking internal state accurately in systems with chat interfaces~\cite{goo_slot-gated_2018,liu_attention-based_2016,lalitha_devi_intent_2023}.
In contrast with ``open'' intent detection systems that attempt to identify \textit{any} intent~\cite{lin_deep_2019,zhang_deep_2021}, the structured nature of ITSs suggests a focus on a limited set of intents or intent categories~\cite{albornoz-de_luise_using_2023}.
The detection of specific intents is tightly coupled to the educational context~\cite{tran_learning_2023,okur_assessing_2023}; a one-size-fits-all approach is unlikely to be effective or to align with the internal state tracked by a specific ITS.
Intent detection is related to affect detection~\cite{dmello_affect_2006}, but is focused on detecting states that correspond to specific system actions.

In this paper, we describe the practical development of an intent detection system within an existing chatbot ITS. 
We compare four different modeling approaches, identifying trade-offs between intent detection accuracy and inference speed, both critical metrics for real-time chat interfaces.
We argue that implementing intent detection in chat-based ITSs is critical for reducing student frustration and that crafting automated conversational experiences with LLMs requires pedagogical choices about when and how students can navigate within an ITS.

\section{Educational Context \& Data}

Digital Harbor Foundation formed a research collaboration with the developers of Rori, an intelligent tutoring system for middle-school math learning.\footnote{\url{https://rori.ai}}
Rori is an English-language chatbot accessed via WhatsApp that is predominantly used in West Africa~\cite{henkel_effective_2024}.
In 2024, Rori began presenting new student users with conversations focused on building meta-cognitive skills and confidence before math practice begins~\cite{levonian_designing_2025}.
In summer 2024, all new students participated in a conversation with the GPT-3.5 large language model about the history of mathematics in Africa.
The conversation uses the highly-structured design of the Rori conversations described by Levonian et al.~\cite{levonian_designing_2025} but is focused on West African math history rather than cultivating a growth mindset.
Table~\ref{tab:maths-history-conversation} shows the phases of the conversation.
The conversation structure and content was designed by educators on the Rori team using culturally relevant pedagogy approaches~\cite{okyere_culturally_2022,gay_culturally_2010}.

In July and August 2024, students completed this math history conversation 2,101 times, sending 10,504 messages.
This conversation presents a pedagogical design problem: what should the system do when a student says they want to do something else or expresses frustration with the current conversation?
The LLM prompt tries to guide students with disengaged or slightly off-topic responses back to the topic at hand, and a moderation system responds to more serious negative content such as violent messages~\cite{levonian_designing_2025}.
A challenging middle ground is messages that indicate a desire to change the topic of the conversations.
How should the system respond in a way that is pedagogically appropriate~\cite{skiba_teaching_2016}?
We approach this as a binary classification problem that can be learned from annotations produced by educators. We train machine learning classifiers to distinguish between messages that should be handled by the LLM and messages that indicate an intent to Change Topic.
Messages with the Change Topic intent are routed not to the next phase of the conversation, but treated as an opportunity to navigate the student to a new topic---a fundamentally different module within the ITS.

\begin{table}[tb]
\centering
\caption{Math history conversation phases and example on-topic messages. Descriptions are representative of the content presented to students at each phase, but they will vary based on a student's responses and questions.}
\label{tab:maths-history-conversation}
\begin{tabularx}{\textwidth} { 
  l
  >{\hsize=1.6\hsize\raggedright\arraybackslash}X 
  >{\hsize=0.4\hsize\raggedright\arraybackslash}X }
\toprule
 & Phase Description & \begin{tabular}[c]{@{}c@{}}Example\\ Responses \end{tabular} \\
\midrule
1 & We will learn about the history of mathematics in this lesson, with a focus on African civilizations. Interesting, isn't it? & yes, yh, okay \\ \midrule
2 & How old do you think maths is? & 100yrs, forever \\ \midrule
3 & Did you know ancient African civilizations played an important part in the development of mathematics? & no, yes, how \\ \midrule
4 & The Mali empire in modern West Africa advanced the knowledge of mathematics through its University of Sankore in Timbuktu. Would you have liked to study there? & yes, maybe, one day, where is that? \\ \midrule
5 & The Bamana Code---developed and used in Africa historically---is the foundation of digital computers. Do you find that exciting? & yes, no, tell me more, what is that \\ \midrule
6 & Are you proud of your African heritage and contributions to the field of math now? & yes, no \\
\bottomrule
\end{tabularx}
\end{table}

We randomly sampled 806 student messages sent during the math history conversation between 1 July 2024 and 28 August 2024.
These messages occur in 208 unique conversations and comprise 7.7\% of student messages sent in that period.
157 students in the annotated sample self-reported a country of origin---82\% said they were from either Nigeria or Ghana.
141 students self-reported an age---median 19 years old, with 30\% of students under the age of 18.

Two of the authors mutually annotated 224 messages, finding acceptable agreement (94.6\% agreement, Cohen's $\kappa$=0.68).
The annotators met to resolve disagreements and then independently annotated the remaining messages.
69 (8.56\%) of annotated messages indicated an early exit.
Qualitatively, many of these messages expressed explicit frustration with the system.
Other messages annotated with the Change Topic intent don't show frustration, e.g. ``Pls can we go straight to the teaching'', ``teach me mathematics'', ``What is the LCM of 16'', ``I want to stop'', ``Can u solve math question 4 me''.

\section{Methods}

We evaluated four modeling approaches for the binary intent classification task.
Annotated data was randomly split into training (60\%), validation (20\%) and test (20\%) sets. 
All hyperparameter tuning was conducted with the validation set and all results are reported on the test set.
We release all analysis and fine-tuning code as well as LLM prompts used.\footnote{\url{https://github.com/DigitalHarborFoundation/its-intent-detection}}
Models are compared by their classification performance (average F1, precision, and recall) and the inference time for processing a single message (in seconds).
Inference time is an important component of a real-time ITS user's quality of experience: slow responses can lead to disengagement.

Our first modeling approach is a conventional, non-LLM machine learning classifier.
We used scikit-learn's random forest implementation with TF-IDF as a strong and fast baseline~\cite{pedregosa_scikit-learn_2011}. We tuned hyperparameters on the validation set using Hyperopt~\cite{bergstra_making_2013}.

Our second modeling approach is a self-hosted LLM fine-tuned for the classification task.
We used the Llama~2 model with 70 billion parameters already fine-tuned for chat use cases~\cite{touvron_llama_2023}.
We fine-tuned Llama~2 for 20 epochs on a single NVIDIA T4 GPU with 16GB of memory using PEFT's implementation of LoRA~\cite{mangrulkar_peft_2022,hu_lora_2022}.
We set fine-tuning hyperparameters based on existing open-source implementations~\cite{talebi_fine-tuning_2023}.
We train a classifier on top of the fine-tuned LoRA model until loss convergence.

Our third modeling approach is to use LLMs hosted via a third-party API.
Use of an API is convenient and enables access to models that may exceed the capabilities of open-source LLMs available for self-hosting, but use of an API call will increase inference time.
We evaluated with OpenAI's GPT-3.5, GPT-4o, GPT-4o Mini, and o3-mini models.
As the conversational context requires a response if the Change Topic intent is not detected, we embed the intent classification problem in the prompt: ``If the user indicates that they want to exit the conversation or change the topic, reply with the text <exit> instead of responding to the student.''
We did minimal prompt engineering.

Our fourth modeling approach is the use of function calling with OpenAI's LLMs.
Function calling involves fine-tuning an LLM to output that a function should be called in certain circumstances rather than generating the expected conversational response~\cite{eleti_function_2023}.
In this case, a function description for a Change Topic request can be provided to the LLM as an alternative to a separate intent classification step or a pre-specified text.\footnote{Function description: ``This function logs that the user wishes to change the topic of conversation from the current discussion to something unrelated.''}
We evaluate the use of function calling for all OpenAI models.
Function calling has previously been used for dialogue state tracking~\cite{li_large_2024}.

\section{Results}


\begin{table}[tb]
\centering
\caption{Performance of various modeling approaches. The right-most columns show precision and recall for the Change Topic intent.}
\label{tab:results}
\begin{tabular}{@{}lcccc|cc@{}}
\toprule
Model & F1 & Precision & Recall & \begin{tabular}[c]{@{}c@{}}Inference\\ Time (s) \end{tabular} & \begin{tabular}[c]{@{}c@{}}Precision\\ (Change) \end{tabular} & \begin{tabular}[c]{@{}c@{}}Recall\\ (Change) \end{tabular}\\
\midrule
Random Forest & 0.62 & 0.64 & 0.60 & \textbf{0.01} & 0.56 & 0.42\\
Llama 2 70B (Fine-tuned) & 0.66 & 0.62 & 0.77 & 0.04 & 0.25 & 0.60 \\
GPT-3.5 & 0.61 & 0.72 & 0.58 & 0.86 & 0.50 & 0.17 \\
GPT-3.5 (Function calling) & 0.57 & 0.60 & 0.84 & 0.83 & 0.20 & 1.00 \\
GPT-4o Mini & 0.71 & 0.81 & 0.66 & 0.94 & 0.67 & 0.33 \\
GPT-4o Mini (Function calling) & 0.73 & 0.79 & 0.70 & 0.98 & 0.62 & 0.42 \\
GPT-4o & \textbf{0.77} & 0.81 & 0.74 & 1.11 & 0.67 & 0.50 \\
GPT-4o (Function calling) & 0.74 & 0.75 & 0.73 & 1.02 & 0.55 & 0.50 \\
o3-mini & 0.67 & 0.70 & 0.65 & 4.99 & 0.44 & 0.33 \\
o3-mini (Function calling) & 0.65 & 0.72 & 0.61 & 4.59 & 0.50 & 0.25 \\
\bottomrule
\end{tabular}
\end{table}

Table~\ref{tab:results} shows performance and inference speed using different models.
A well-tuned conventional machine learning approach achieves a 0.62 F1 score, indicating that the underlying classification task is challenging.
Fine-tuned Llama~2 improves F1 score to 0.66, at the cost of an increased per-message inference time from less than 0.01 seconds to 0.035 seconds on average.

Among the OpenAI-hosted LLMs, GPT-3.5 fails to outperform the baseline. However, all other models outperform fine-tuned Llama~2, with GPT-4o outperforming GPT-4o Mini as expected of the model size.
All OpenAI models take around 1 second to retrieve a classification, with the exception of the reasoning model o3-mini which offers both slower response times and worse performance on this task.
The function calling approach performs similarly to the basic prompting approach, suggesting that either approach is reasonable for this task.

\section{Discussion}

Intent detection is a critical modeling task for ITSs that incorporate LLMs into chat-like interfaces.
We find only modest performance on the specific binary classification task we investigated here; any of these models will produce many misclassifications that will need to be gracefully handled by the ITS.
As LLMs proliferate within ITSs, the use of iterative dialogue flows for confirmation and navigation are an important focus area for future design work.
Within Rori, we intend to handle messages differently based on the intent detection classifier's confidence, handling high-confidence messages without clarification but presenting confirmatory dialogue turns when classifier confidence is low.
This is a downside to the use of third-party LLMs for intent detection problems: their relative lack of calibrated confidence estimates.
Incorporating research on probing models for their certainty~\cite{jiang_how_2021,geng_survey_2024} might enable the wider use of LLM-based intent detection models.
During internal development, access to calibrated probability estimates also enables the use of model comparison metrics more appropriate for dialogue classification tasks, such as ROC AUC~\cite{lin_robust_2023}.

The binary intent detection task we considered here is fundamentally limited; conversational interfaces require designing for the flexibility in human conversations~\cite{xu_are_2021}.
Xu et al. characterize existing conversational user interfaces as providing mere ``psuedo-conversations'', as systems lack the technical infrastructure for collaborative conversation~\cite{xu_are_2021}.
In our view, the same critique applies to LLM-backed ITSs:
conducting meaningful conversational lessons requires both detecting a wide variety of intents and providing an appropriate system response to each detected intent.
In other words, knowing that a student wants to change topics within an ITS is only the beginning of designing navigational capabilities that align with the pedagogical guardrails that human tutors apply instinctively.
Chat interfaces should demonstrate the extent to which they can facilitate various student intents by making some internal state visible, avoiding needless anthropomorphization~\cite{gabriel_ethics_2024}, and teaching student users about what capabilities are provided. 
In other cases, chat interfaces can be supplemented or replaced entirely with more traditional interfaces, which excel at presenting a uniform view of information and responding predictably to users' inputs.

Our modeling results demonstrate trade-offs between inference time---ex\-perienced directly as response latency by users---and intent detection accuracy.
While an increase in response latency is often perceived as negative from a user experience perspective~\cite{shneiderman_response_1984}, the impact of delayed responses in chat interfaces is less clear~\cite{Gnewuch_opposing_2022}.
ITS designers should carefully consider the impact of delaying a response in order to ensure a more appropriate response.

This preliminary investigation has many limitations.
Our test dataset consists of only 162 messages, which makes fine-grained comparison between models imprecise.
Computing standard errors is challenging because messages are clustered within conversations; our future work will use cross validation and clustered standard errors to enable more reliable model comparison~\cite{miller_adding_2024}.
Further, the generalizability of these results is unclear: we studied only a single binary intent detection task.
Future work should investigate a broader set of intents and explore the degree to which particular intent models can be applied in other ITSs~\cite{albornoz-de_luise_using_2023}.
Intent detection is an important problem because it directly impacts user experience, but we did not collect any user experience metrics in this study.
Modeling results should be contextualized through usability studies that investigate the impact of these models on student use when these intent detection systems are deployed in the wild. 


\begin{credits}
\subsubsection{\ackname}
We would like to thank Owen Henkel, Millie-Ellen Postle, Maria Dyshel, Greg Thompson, Bill Roberts, and the staff of Rising Academies.
This work was supported by the Learning Engineering Virtual Institute (LEVI) and by the Digital Harbor Foundation.

\subsubsection{\discintname}
The authors have no competing interests to declare that are relevant to the content of this article. No Generative AI technologies were used in the drafting of this article.
\end{credits}
%
%
%
\bibliographystyle{splncs04}
\bibliography{levi_enghub}

\end{document}